\UseRawInputEncoding 
\documentclass[aps,prb,twocolumn,groupedaddress,showpacs]{revtex4}

\usepackage{color}
\usepackage{amssymb}
\usepackage{graphicx}
\usepackage{amsmath}
\usepackage{dcolumn}
\usepackage{bm}
\usepackage{textgreek}

\usepackage[colorlinks=true,linkcolor=blue,citecolor=blue,urlcolor=black]{hyperref}%
\expandafter\ifx\csname package@font\endcsname\relax\else
 \expandafter\expandafter
 \expandafter\usepackage
\expandafter\expandafter
 \expandafter{\csname package@font\endcsname}%
\fi

\begin{document}

\title{ٰVertical Strain-Induced Modification of the Electrical and Spin Properties of Monolayer MoSi$_2$X$_4$ (X= N, P, As and Sb)}

\author{Shoeib Babaee Touski}
\affiliation{Department of Electrical Engineering, Hamedan University of Technology, Hamedan, Iran}
\author{Nayereh Ghobadi}
\email{n.ghobadi@znu.ac.ir}
\affiliation{Department of Electrical Engineering, University of Zanjan, Zanjan, Iran}

\date{\today}

\begin{abstract}
In this work, the electrical and spin properties of monolayer MoSi$_2$X$_4$ (X= N, P, As, and Sb) under vertical strain are investigated. The band structures state that MoSi$_2$N$_4$ is an indirect semiconductor, whereas other compounds are direct semiconductors. The vertical strain has been selected to modify the electrical properties. The bandgap shows a maximum and decreases for both tensile and compressive strains. The valence band at K-point displays a large spin-splitting, whereas the conduction band has a negligible splitting. On the other hand, the second conduction band has a large spin-splitting and moves down under vertical strain which leads to a large spin-splitting in both conduction and valence bands edges. The projected density of states along with the projected band structure clarifies the origin of these large spin-splittings. These three spin-splittings can be controlled by vertical strain. 
\end{abstract} 

%\pacs{78.67.Lt, 73.22.-f, 78.20.Bh}

\maketitle
%%%%%%%%%%%%%%%%%%%%%%%%%%%%%%%%%%%%%%%%%%%%%%%%%%%%%%%%%%%%%%%%%%%%%%%%%%%%%%%%%%%%%%%
%%%%%%%%%%%%%%%%%%%% I N T R O D U C T I O N %%%%%%%%%%%%%%%%%%%%%%%%%%%%%%%%%%%%%%%%%%
%%%%%%%%%%%%%%%%%%%%%%%%%%%%%%%%%%%%%%%%%%%%%%%%%%%%%%%%%%%%%%%%%%%%%%%%%%%%%%%%%%%%%%%
\section{Introduction}
\label{intro}
The graphene, the first member of two-dimensional (2D) materials, with honeycomb structure does not exhibit an electronic band gap \cite{neto2009electronic}. This problem has motivated the researchers to explore and design novel two-dimensional semiconductors, such as the transition metal dichalcogenides \cite{radisavljevic2011single}, black phosphorene \cite{guan2014phase}, antimonene \cite{zhang2015atomically} and indium selenide\cite{mudd2013tuning}. Therefore, exploring new 2D materials with proper electronic properties for particular applications is highly demanded.

After monolayer TMDC, 2D transition metal nitride (TMN) have been proposed in the recent years \cite{li2017computational,wang2016strain,ozdemir2018intercalation}. Since TMNs do not contain layered structures, the fabrication of large-area TMNs monolayers has remained challenging. Recently, MoSi$_2$N$_4$ monolayer based on TMN and without 3D counterpart structure has been synthesized\cite{hong2020chemical}. MoSi$_2$N$_4$ is constructed from a MoN$_2$ monolayer sandwiched between two Si-N monolayers. By analyzing MoSi$_2$N$_4$ monolayer, it can be realized that with sandwiching a TMDC-type MZ$_2$ monolayer into InSe-type A$_2$Z$_2$, twelve different structures with MA$_2$Z$_4$ formula have been achieved. They demonstrate different phases such as semiconductor, topological insulators, and Ising superconductors \cite{wang2021}.
Monolayer MoSi$_2$N$_4$ has been successfully fabricated using chemical vapor deposition with large size of up to 15mm$\times$15mm\cite{hong2020chemical}. This compound demonstrates high strength and excellent stability at ambient. MoSi$_2$N$_4$ compound also shows an elastic constant three times larger than monolayer MoS$_2$. In addition, the large electron and hole mobility is estimated to be 270/1200 cm$^2$/Vs that is four to six times higher than MoS$_2$ ones. This high mobility along with outstanding stability in the ambient environment makes this material a promising candidate for future applications. 

Other members of MoSi$_2$N$_4$ family with MA$_2$Z$_4$ formula (M = Transition metal, A = Si, Ge and Z = N, P, As) have been extensively investigated \cite{zhong2021strain,wang2021}. These monolayers exhibit a wide range from semiconducting to metallic properties. In addition, some compounds with magnetic transition metal elements also represent magnetic properties\cite{li2021design}. The MoSi$_2$N$_4$ and WSi$_2$N$_4$ monolayers also show high lattice thermal conductivity for thermoelectric applications \cite{mortazavi2021exceptional,yu2021high}. 
Monolayer MoSi$_2$N$_4$ and its family also have suitable band gaps up to 1.73eV for potential optical applications in the visible range\cite{bafekry2021mosi2n4,yao2021novel}.

It has been anticipated that MoSi$_2$N$_4$ family compounds have a pair of valley pseudospins similar to monolayer MoS$_2$ \cite{yang2021valley}. These monolayers have been also predicted to exceed monolayer MoS$_2$ due to fascinating valleytronic properties, for instance, multiple-folded valleys in monolayer MoSi$_2$As$_4$. 
The lack of inversion symmetry along with strong spin-orbit coupling from transition metal elements in the MA$_2$Z$_4$ family results in two inequivalent valleys (K and K') with sizable spin-splitting and spin-valley coupling \cite{ai2021theoretical}. For example, WSi$_2$N$_4$ demonstrates a large spin-splitting at both conduction and valence bands at K and K'-valleys together with spin-valley coupling.

It has been confirmed theoretically and experimentally that a semiconductor to metal transition occurs in multi-layer MoS$_2$ by applying vertical strain \cite{nayak2014pressure,chi2014pressure,ghobadi2019normal}. Totally, the decrease of interlayer distance may lead to charge redistribution and semiconductor to metal transition. The out-of-plane strain is presented as a powerful tool to tune the electrical properties of bilayer MA$_2$Z$_4$ family \cite{bafekry2021effect}. For instance, the bandgap of bilayer MoSi$_2$N$_4$ decreases with increasing compressive strain and the bandgap closes at the strain of 22$\%$. Such a semiconductor to metal transition also occurs in other MA$_2$Z$_4$ bilayers \cite{zhong2021strain}. The transition pressure is distributed from 2.18 GPa in CrSi$_2$N$_4$ to 32.04 GPa in TiSi$_2$N$_4$. In another work, the biaxial strain is used to tune the band gaps of bilayer MoSi$_2$N$_4$ and WSi$_2$N$_4$ \cite{wu2021semiconductor}. It has been reported that these compounds demonstrate a direct bandgap at compressive strain. Furthermore, it has been shown that strain can also modify the electronic and magnetic properties of VSi$_2$P$_4$ monolayer \cite{guo2020coexistence}. The strain increasing changes the phase of this monolayer from a ferromagnetic metal to a spin-gapless semiconductor, afterward to a ferromagnetic semiconductor, and then come back to spin-gapless semiconductor and finally to a ferromagnetic half-metal. 

In this work, the effects of vertical strain on the electrical and spin properties of monolayer MoSi$_2$X$_4$ (X=N, P,  As, and Sb) are investigated. First, the electrical properties such as bandgap, effective mass, and band edge locations are obtained at equilibrium. In the following, vertical strain is applied to all materials, and the variation of band gaps, band edges, and charge densities are explored. Finally, spin-splitting at both equilibrium and strained samples is studied.

\section{Computational details}
In order to investigate the electrical and spin properties of MoSi$_2$X$_4$ (X= N, P, As and Sb) monolayers, density functional calculations are performed using the SIESTA package \cite{soler2002siesta}. The generalized gradient approximation (GGA) with the Perdew-Burke-Ernzerhof (PBE) \cite{perdew1981self} functional is employed for the exchange-correlation term. We have adopted fully relativistic pseudopotentials and have taken into account the effect of spin-orbit coupling (SOC). A Monkhorst-Pack k-point grid of $21\times21\times1$ is chosen for the unit-cell. The energy cutoff is set to be 200 Ry and a double-$\zeta$ plus polarization basis-set is used. The total energy is converged to better than $10^{-5}$ eV and the geometries are fully relaxed until the force on each atom is less than 0.02 eV/$\mathrm{\AA}$. A vacuum region of 30 $\mathrm{\AA}$ is added to avoid interactions in the normal direction. To visualize the atomic structures, XCrySDen package has been used \cite{kokalj2003computer}. The out-of-plane strain is defined as $\varepsilon=(t-t_0)/t_0$, where $t_0$ and $t$ are the equilibrium and deformed compound thickness, respectively. The effective masses are calculated by using the following equation \cite{touski2020interplay,ghobadi2020electrical},
\begin{equation}
m^*=\hbar^2/\left(\partial^2E/\partial k^2\right)
\end{equation}
Here, $\hbar$ is the reduced Planck constant, E and k are the energy and wave vector of conduction band minimum and valence band maximum.

%%%%%%%%%%%%%%%%%%%%%%%%%%%%%%%%%%%%%%%%%%%%%%%%%%%%%%%%%%%%%%%%%%%%%%%%%%%%%%%%%%%%%%%
%%%%%%%%%%%%%%%%%%%%%%%%     NUMERICAL RESULTS        %%%%%%%%%%%%%%%%%%%%%%%%%%%%%%%%
%%%%%%%%%%%%%%%%%%%%%%%%%%%%%%%%%%%%%%%%%%%%%%%%%%%%%%%%%%%%%%%%%%%%%%%%%%%%%%%%%%%%%%%

\begin{figure}
	\centering
	\includegraphics[width=1.0\linewidth]{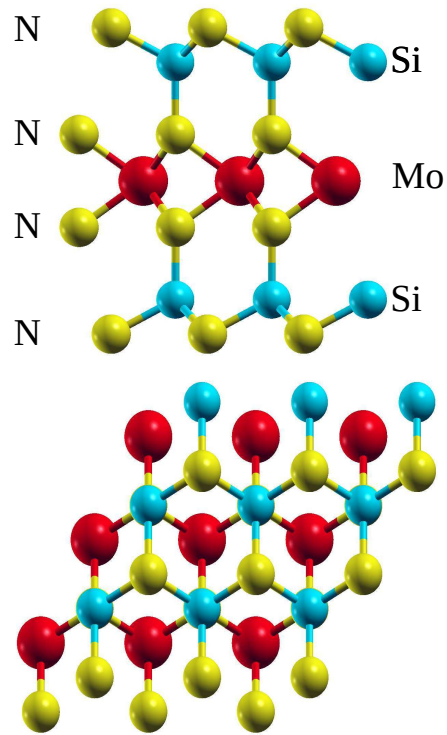}
	\caption{The schematic of the studied materials with MoSi$_2$X$_4$ formula. The atomic structure of MoSi$_2$N$_4$ is depicted. N atoms are replaced by P, As and Sb in other studied structures.}
	\label{fig:fig1}
\end{figure}

\begin{figure*}
	\centering
	\includegraphics[width=1.0\linewidth]{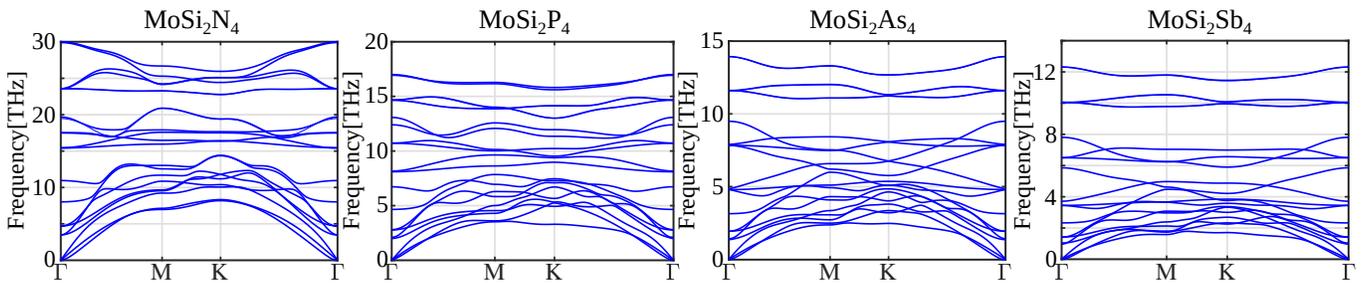}
	\caption{Phonon dispersion relations for four compounds.}
	\label{fig:phonon}
\end{figure*}

\begin{figure*}
	\centering
	\includegraphics[width=1.0\linewidth]{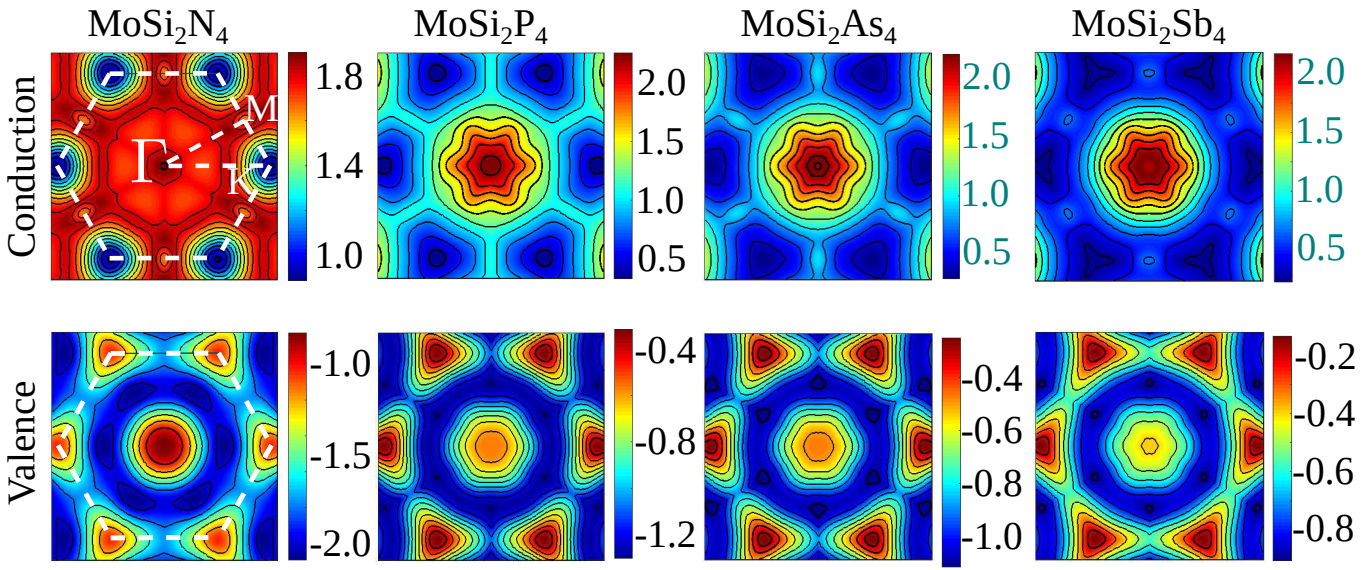}
	\caption{Two-dimensional map of the conduction and valence bands for all compounds. The first Brillouin zone is indicated for MoSi$_2$N$_4$ as a sample. Top and down rows are for the first conduction and valence bands, respectively.}
	\label{fig:con-val}
\end{figure*}

\begin{table*}[t]
	\caption{ The lattice constant (a), the Mo-X (d$_{Mo-X}$) and Si-X (d$_{Si-X}$) bond lengths, the vertical distance between Si and X atoms ($\Delta_{Si-X}$), the thickness and the elastic constants (C$_{11}$ and C$_{12}$) of MoSi$_2$X$_4$ monolayers. \label{tab:tab1}}
	
	\begin{tabular}{p{1.5cm}p{1.7cm}p{1.7cm}p{1.7cm}p{1.7cm}p{2cm}p{1.7cm}p{1.7cm}}
		\hline
		\hline
		&a($\mathrm{\AA}$) & d$_{Mo-X}$($\mathrm{\AA}$)  & d$_{Si-X}$($\mathrm{\AA}$) & $\Delta_{Si-X}$($\mathrm{\AA}$) & Thickness($\mathrm{\AA}$) & C$_{11}$(N/m) & C$_{12}$(N/m) \\
		\hline
		MoSi$_2$N$_4$ &2.928  &2.111  & 1.763 &0.531  & 7.119 & 472.66 & 142.02    \\
		
		\hline
		MoSi$_2$P$_4$ &3.486  &2.477  & 2.253 & 1.046 &9.486  & 207.77 & 58.46 \\
		
		\hline
		MoSi$_2$As$_4$ &3.633  & 2.583 & 2.366 & 1.136 & 10.021 & 178.03 & 54.75\\
		
		\hline
		MoSi$_2$Sb$_4$ & 3.898 &2.777  &2.589  &1.278  &10.991  & 141.40 &47.72  \\

		\hline
		\hline
	\end{tabular}
\end{table*}

\section{Results and discussion}
\label{Result} 
The schematic of the studied compounds is shown in Fig. \ref{fig:fig1}. As it is obvious, a MoN$_2$ monolayer is sandwiched between two SiN monolayers. The unsaturated Si atoms in SiN sub-layer bond with unsaturated N atoms in MoN$_2$. The structural and mechanical properties of four monolayers are presented in Table \ref{tab:tab1}. The lattice constant of MoSi$_2$N$_4$ is 2.928 $\AA$ that is close to experimental result\cite{hong2020chemical}. The lattice constants of MoSi$_2$P$_4$ and MoSi$_2$As$_4$ are also 3.486 and 3.633 $\AA$, respectively, that are compatible with previous studies \cite{guo2020coexistence,ai2021theoretical}. The lattice constant increases for the compounds with heavier X elements. This behavior can be observed in the Mo-X bond length ($d_{Mo-X}$) in MoX$_2$ sub-layer, Si-X bond length, and buckling height of SiN sub-layer.  Thickness of these compounds demonstrates a high value  of 7.119 for MoSi$_2$N$_4$ to 10.991 $\AA$ for MoSi$_2$Sb$_4$. This high thickness confirms that applying vertical strain is feasible for these compounds. The elastic constants C$_{11}$ and C$_{12}$ are also obtained and reported in the table which is compatible with previously reported amounts  \cite{yao2021novel}. These materials are stable and the elastic constants satisfy the Born criteria stability, $0<C_{11}$, $0<C_{22}$, $C_{12}<C_{11}$ and $C_{12}<C_{22}$. In addition, the stability of these compounds is studied by phonon dispersion, see Fig. \ref{fig:phonon}. The positive frequencies confirm the stability of these structures. 

The two-dimensional contour plots of the first Brillouin zone for the valence and conduction bands are depicted in Fig. \ref{fig:con-val}. The conduction band minimum (CBM) is located at K-valley for all compounds. In the case of MoSi$_2$N$_4$, K-valley approximately demonstrates isotropic contour in different directions whereas, anisotropic behavior is obvious for the other compounds. The anisotropy of K-valley increases for heavier compounds and MoSi$_2$Sb$_4$ demonstrates the highest anisotropy. At the same time, the valence band maximum (VBM) is located at $\Gamma$-point in the case of MoSi$_2$N$_4$, whereas the energy of K-point is close to the $\Gamma$-point. On the other hand, the VBM of other structures is placed at K-point. MoSi$_2$N$_4$ demonstrates isotropic contours around $\Gamma$-point whereas the other compounds exhibit anisotropic behavior. However, K-point in the valence band shows a high anisotropy with triangular contours for all compounds.

\begin{figure*}
	\centering
	\includegraphics[width=0.9\linewidth]{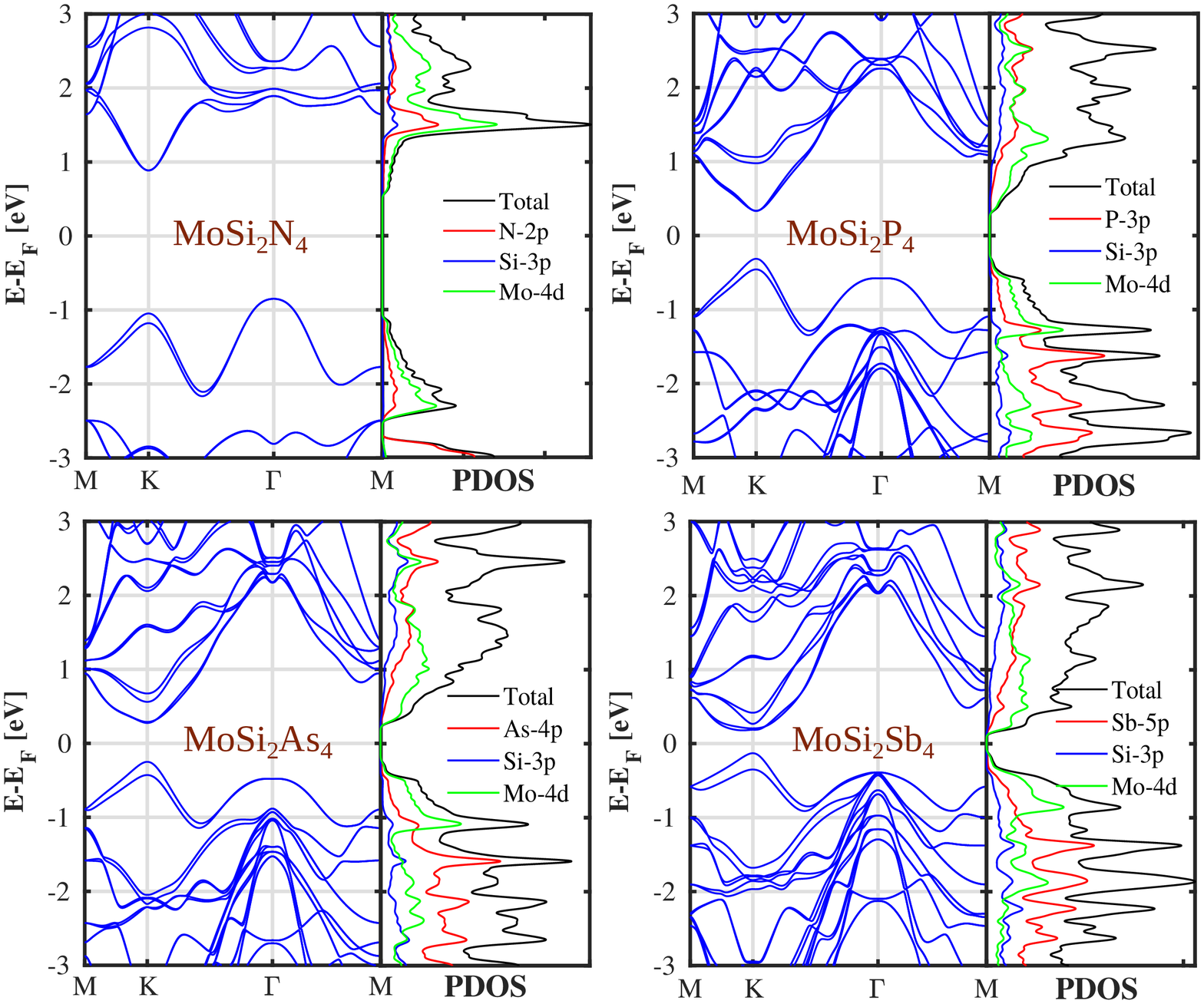}
	\caption{The band structure along with corresponding PDOS of MoSi$_2$X$_4$ structures. The d-orbital of Mo atom and p-orbitals of Si and X atoms have the main contributions to the DOS.}
	\label{fig:fig2}
\end{figure*}

The band structures of the four studied materials are drawn in Fig. \ref{fig:fig2}. MoSi$_2$N$_4$ demonstrates an indirect bandgap from $\Gamma$-valley in the valence to K-valley in the conduction band whereas, K-point at the valence band is close to $\Gamma$-point and contributes to the top of the valence band. Three other compounds have a direct bandgap at K-point. In these compounds, $\Gamma$-point at valence band also is close to K-point. On the other hand, K-point displays a considerable spin-splitting whereas, the splitting vanishes for $\Gamma$-point due to the high symmetry. The minimum of the second conduction band with a large spin-splitting is near to the minimum of the first band. In order to further analyze the electronic properties of the
structures, the projected density of states (PDOS) is also depicted in Fig. \ref{fig:fig2}. The states near the VBM and CBM are mainly contributed by the Mo atom. X atoms also contribute to the conduction and valence bands, but their contributions are lower than the Mo atom.  

\begin{figure}
	\centering
	\includegraphics[width=1.0\linewidth]{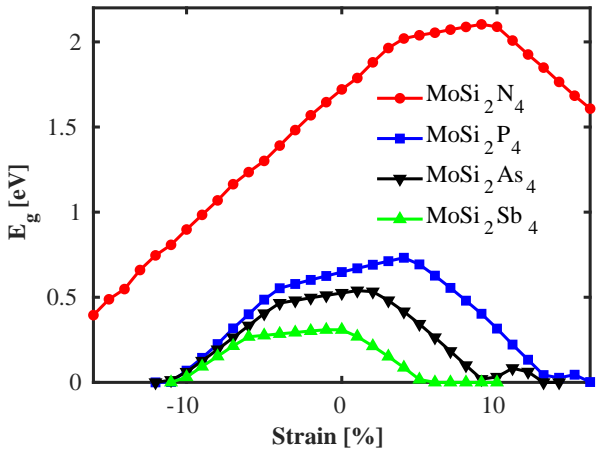}
	\caption{The band gap variation as a function of vertical strain.}
	\label{fig:fig3}
\end{figure}

\begin{figure*}
\centering
\includegraphics[width=1.0\linewidth]{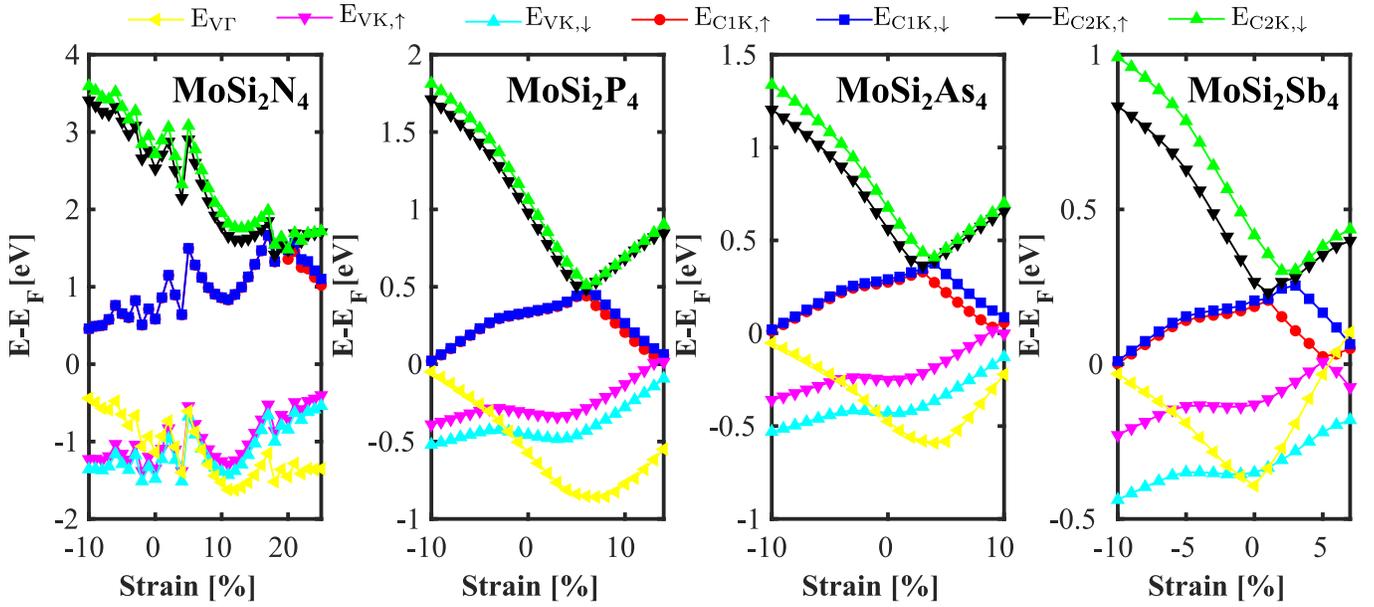}
\caption{The energies of the K-valleys of the first and second conduction bands along with K- and $\Gamma$-points of the valence band versus vertical strain. The up- and down-spin for each band are also indicated. The range of strain is kept in the semiconducting phase and is different for various compounds. }
\label{fig:energy}
\end{figure*}

\begin{figure*}
	\centering
	\includegraphics[width=1.0\linewidth]{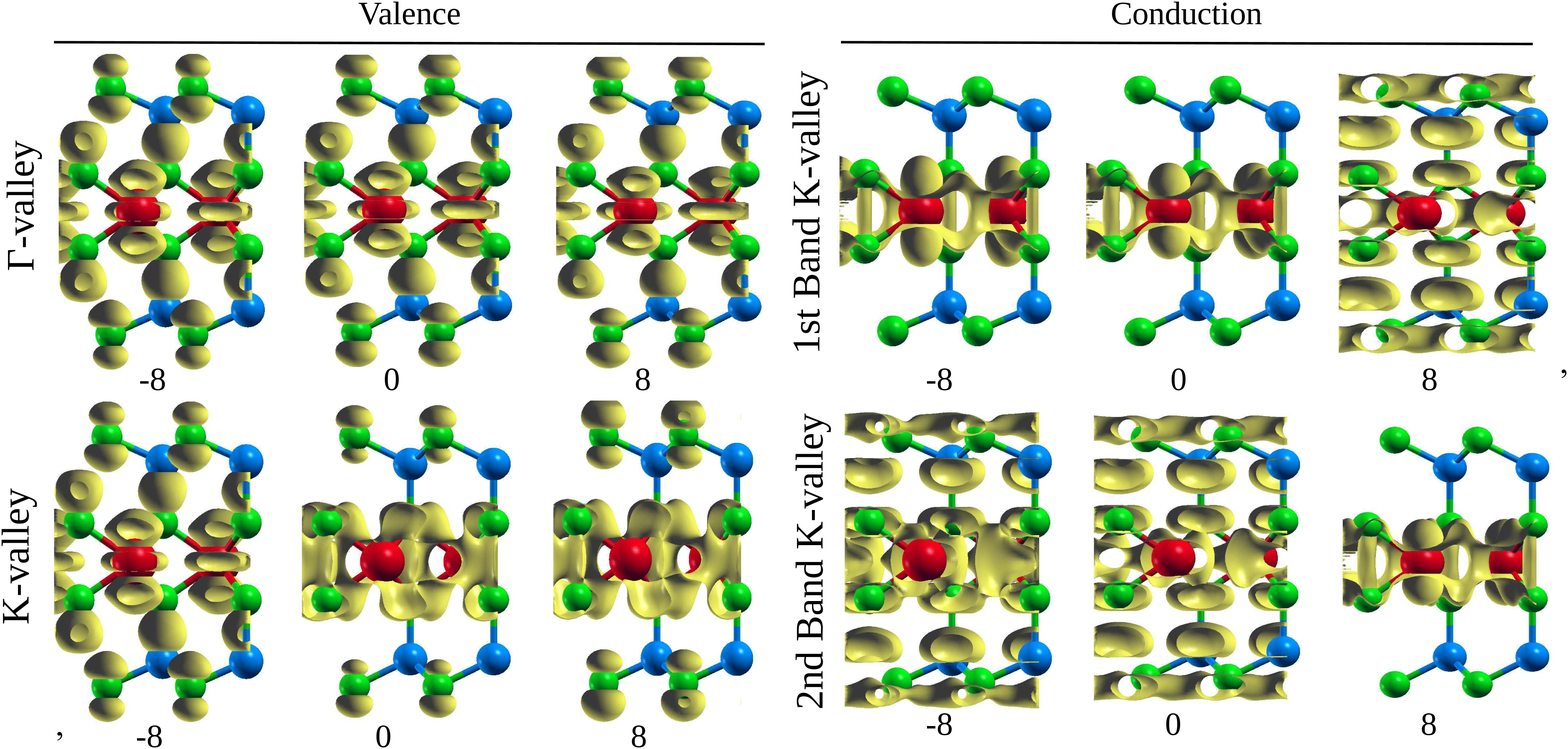}
	\caption{Charge distribution for MoSi$_2$P$_4$ as a sample. The charge is plotted for \textGamma- and K-point in the valence band and for the first and second conduction band at K-valley.}
	\label{fig:charge}
\end{figure*}

The electrical and spin properties of four materials are summarized in Table \ref{tab:tab2}. First, the band gaps with and without spin-orbit coupling consideration are compared. In the case of MoSi$_2$N$_4$, the band gaps without and with SOC are approximately the same. These values which are in good agreement with experimental work \cite{hong2020chemical}, are 1.73 and 1.721 eV, respectively. The bandgap of MoSi$_2$P$_4$ without(with) SOC is 0.772(0.648) eV that is near to 0.61 eV reported for with SOC\cite{ai2021theoretical}. The bandgap of MoSi$_2$As$_4$ decreases from 0.664 to 0.525eV with applying spin-orbit coupling, while the values of 0.6eV and 0.41-0.5 eV are reported for without and with spin-orbit coupling consideration, respectively\cite{hong2020chemical,guo2020coexistence,ai2021theoretical,li2020valley}. The SOC has more significant effects on the heavier compounds so that the bandgap reduces by 0.172 eV for MoSi$_2$Sb$_4$ with applying SOC. In addition, the conduction and valence band maximum is presented in the table. In the end, the spin-splitting at K-point in the valence band, the first and second conduction bands are listed in the table. The value of spin-splitting of MoSi$_2$N$_4$ at K-point ($\lambda_{K,V}$) of the valence band is 131 meV which is a little smaller than 140 meV from experimental work. However, 130 meV is reported for $\lambda_{K,V}$ in the theoretical papers \cite{li2020valley}. Furthermore, the calculated value of $\lambda_{K,V}$ for MoSi$_2$As$_4$ is 0.179 eV and in good agreement with previous reported values\cite{guo2020coexistence,li2020valley}. $\lambda_{K,V}$ increases with changing X-atom from N to Sb and reaches 220 meV for MoSi$_2$Sb$_4$. The spin-splitting at the first conduction band ($\lambda_{K,C1}$) is not considerable. On the contrary, the second band demonstrates a significant spin-splitting and for example, is larger than $\lambda_{K,V}$ in MoSi$_2$N$_4$, but the energy of the second band is much higher than the first band in this compound. The spin-splitting at K-valley in the second conduction band ($\lambda_{K,C2}$) decreases from 187 meV in MoSi$_2$N$_4$ to 83 meV in MoSi$_2$P$_4$ and increases by changing X atom from P to Sb. By looking at the conduction band of MoSi$_2$As$_4$ and MoSi$_2$Sb$_4$ at K-point, one can find that the first and third bands do not demonstrate remarkable spin-splitting, while the second and fourth display notable spin-splitting.

\begin{figure*}
	\centering
	\includegraphics[width=1.0\linewidth]{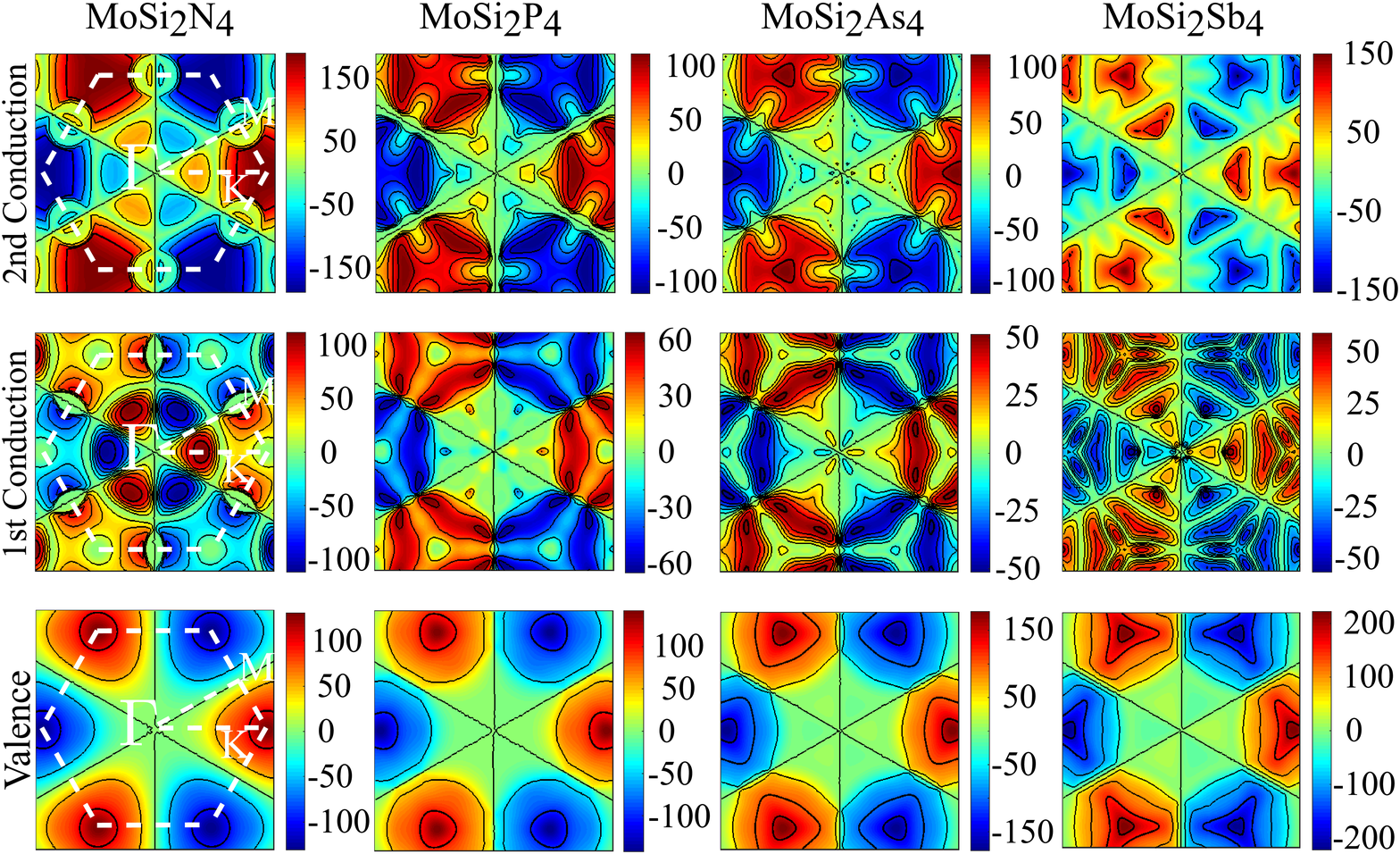}
	\caption{Two-dimensional map of the spin-splitting in the first and second conduction bands along with valence band. The first Brillouin zone is highlighted by the dashed line in the first compound. The large spin-splitting at valence and second conduction band is obvious.}
	\label{fig:ssc-ssv}
\end{figure*}

\begin{table*}[t]
	\caption{The band gap calculated with PBE functional (E$^{PBE}_g$) and PBE with spin-orbit coupling included (E$^{SOC}_g$), the energies of conduction (E$_C$-E$_F$) and valence band (E$_V$-E$_F$) edges, the compressive strain ($\varepsilon_{trans}$) and pressure (P$_{trans}$) required for semiconductor to metal transition, the spin-splitting at K point of the valence band ($\lambda_{K_V}$) and first and second bands of conduction ($\lambda_{K_{C1}}$ and $\lambda_{K_{C2}}$). \label{tab:tab2}}
	\begin{tabular}{p{1.5cm}p{1.2cm}p{1.2cm}p{1.2cm}p{1.2cm}p{1.2cm}p{1.2cm}p{1.2cm}p{1.2cm}p{1.2cm}p{1.2cm}p{1.2cm}}
		\hline
		\hline
		& E$^{PBE}_g$ & E$^{SOC}_g$ & VBM & CBM & E$_C$-E$_F$ & E$_V$-E$_F$ &$\varepsilon_{trans}$&$P_{trans}$&  $\lambda_{K_V}$  & $\lambda_{K_{C1}}$  & $\lambda_{K_{C2}}$    \\
		& (eV) & (eV) &  &  & (eV) & (eV)  &($\%$)&(GPa)& (eV) & (eV)  & (eV)  \\
		\hline

		MoSi$_2$N$_4$	&1.73 &1.721 &$\Gamma$ & K& 0.582 &-1.139 &-22&25.3& 0.131 &	0.003&	0.187 \\
		
		\hline
		
		MoSi$_2$P$_4$	&0.772 &0.648 &K &K &0.333 &-0.315 &-11&9.1&0.143&	0.004&	0.083 \\
		
		\hline
		
		MoSi$_2$As$_4$	&0.664 &0.525 &K &K &0.275 &-0.25 &-11&8.9& 0.179&	0.014&	0.113 \\
		
		\hline
		
		MoSi$_2$Sb$_4$	&0.481 &0.309 &K &K &0.179 &-0.131 &-10&7.3& 0.22&	0.018&	0.151 \\
		
		\hline
		\hline
	\end{tabular}
\end{table*}

\begin{table}[t]
	\caption{The effective masses at K- and $\Gamma$-points of the valence band and K-valley in the conduction band. All reported effective masses are in m$_0$ unit.
		\label{tab:tab3}}
	\begin{tabular}{p{1.5cm}p{1.0cm}p{1.0cm}p{1.0cm}p{1.0cm}p{1.0cm}p{1.0cm}}
		\hline
		\hline
		&$m^{v,*}_{K\rightarrow M}$ & $m^{v,*}_{K\rightarrow \Gamma}$ & $m^{v,*}_{\Gamma\rightarrow K}$ & $m^{v,*}_{\Gamma\rightarrow M}$ & $m^{c,*}_{K\rightarrow M}$ & $m^{c,*}_{K\rightarrow \Gamma}$   \\
		\hline
		MoSi$_2$N$_4$	&0.794&	0.595&	1.375&	1.37 &0.544&	0.485    \\
		\hline
		MoSi$_2$P$_4$	&0.562&	0.388&	1.483&	1.335 &0.697&	0.567    \\
		\hline
		MoSi$_2$As$_4$	&0.701&	0.438&	1.586&	1.407 &1.206&	0.625    \\
		\hline
		MoSi$_2$Sb$_4$	&0.925&	0.447&	2.288&	1.992 &1.843&	0.647    \\
		\hline
		\hline
	\end{tabular}
\end{table}

The effective masses of the conduction and valence bands are reported in Table \ref{tab:tab3}. The effective masses of both $\Gamma$- and K-points of the valence band are presented in the table. K-point demonstrates a lower effective mass with respect to $\Gamma$-one. The anisotropic contour around K-valley leads to two different values at K to M and K to $\Gamma$ paths. The K to M effective mass ($m^{V*}_{K\rightarrow M}$) displays a higher value than K to \textGamma~effective mass ($m^{V*}_{K\rightarrow \Gamma}$). On the other hand, \textGamma-point owns isotropic contour, and effective masses at different paths are approximately equal. In Ref. \onlinecite{hong2020chemical}, the effective mass of the electron is reported to be 0.486 m$_0$, which corresponds to our calculated value 0.485 m$_0$ for K to \textGamma~path. Furthermore, the reported value of the hole effective mass is 0.683 m$_0$ that is in the middle of our calculated effective masses at K-valley. However, \textGamma-point is the VBM and demonstrates a larger effective masses.  
Except MoSi$_2$N$_4$, the effective masses for the valence band increase for heavier compounds and MoSi$_2$P$_4$ contains the lowest effective mass. On the other hand, only K-valley contributes to the conduction band. The effective masses at two different paths for this valley are reported in the table. The effective masses in these two paths are in the same range for MoSi$_2$N$_4$ and MoSi$_2$P$_4$, whereas their differences are enhanced for MoSi$_2$As$_4$ and MoSi$_2$Sb$_4$. As mentioned before, the contour around K-valley changes from circular for MoSi$_2$N$_4$ to triangular in MoSi$_2$Sb$_4$. In this regard, MoSi$_2$Sb$_4$ demonstrates high anisotropic effective masses at two different paths. The effective masses of the conduction band increase for the heavier compounds. MoSi$_2$N$_4$ and MoSi$_2$P$_4$ possess the lowest effective mass in the conduction and valence bands, respectively.

The vertical strain is known as a powerful tool to modify electrical and spin properties of two-dimensional compounds \cite{ghobadi2021structural,shamekhi2020band}. This strain can be applied by out-of-plane pressure and 2D materials can be used as a pressure sensor. The vertical strain is applied to the four structures and their electrical and spin properties are studied. First, the variation of the band gaps as a function of vertical strain is depicted in Fig. \ref{fig:fig3}. The band gaps exhibit a maximum value at a specific small tensile strain and then decrease for larger compressive and tensile strains. The band gaps vanish at transition strains ($\varepsilon_{trans}$). The value of compressive $\varepsilon_{trans}$ and its corresponding pressure ($P_{trans}$) are reported in Table \ref{tab:tab2}. The band gaps of MoSi$_2$P$_4$, MoSi$_2$As$_4$ and MoSi$_2$Sb$_4$ vanish around the strain of -10$\%$ or -11$\%$. On the other hand, MoSi$_2$N$_4$ demonstrates a higher bandgap, and $\varepsilon_{trans}$ is two times larger at -22$\%$ strain. 
$P_{trans}$ for MoSi$_2$P$_4$, MoSi$_2$As$_4$ and MoSi$_2$Sb$_4$ is in the range of [7.3-9.1] GPa, whereas MoSi$_2$N$_4$ demonstrates much larger $P_{trans}$ of 25.3 GPa. The transition pressures for bilayers are reported from 2.18 GPa in CrSi$_2$N$_4$ to 32.04 GPa in TiSi$_2$N$_4$ \cite{zhong2021strain} that are comparable with MoSi$_2$X$_4$ monolayer. This transition pressure range also proves that applying out-of-strain to monolayer MoSi$_2$X$_4$ is feasible. The band gaps also vanish at tensile strain. MoSi$_2$Sb$_4$ demonstrates higher sensitivity to tensile strain and its band gap vanishes at lower strains. The transition strain at tensile regime increases as X atom changes from Sb to N. One can conclude from the figure that tensile transition strain of MoSi$_2$N$_4$ is much larger than 15$\%$.

\begin{figure*}
	\centering
	\includegraphics[width=0.8\linewidth]{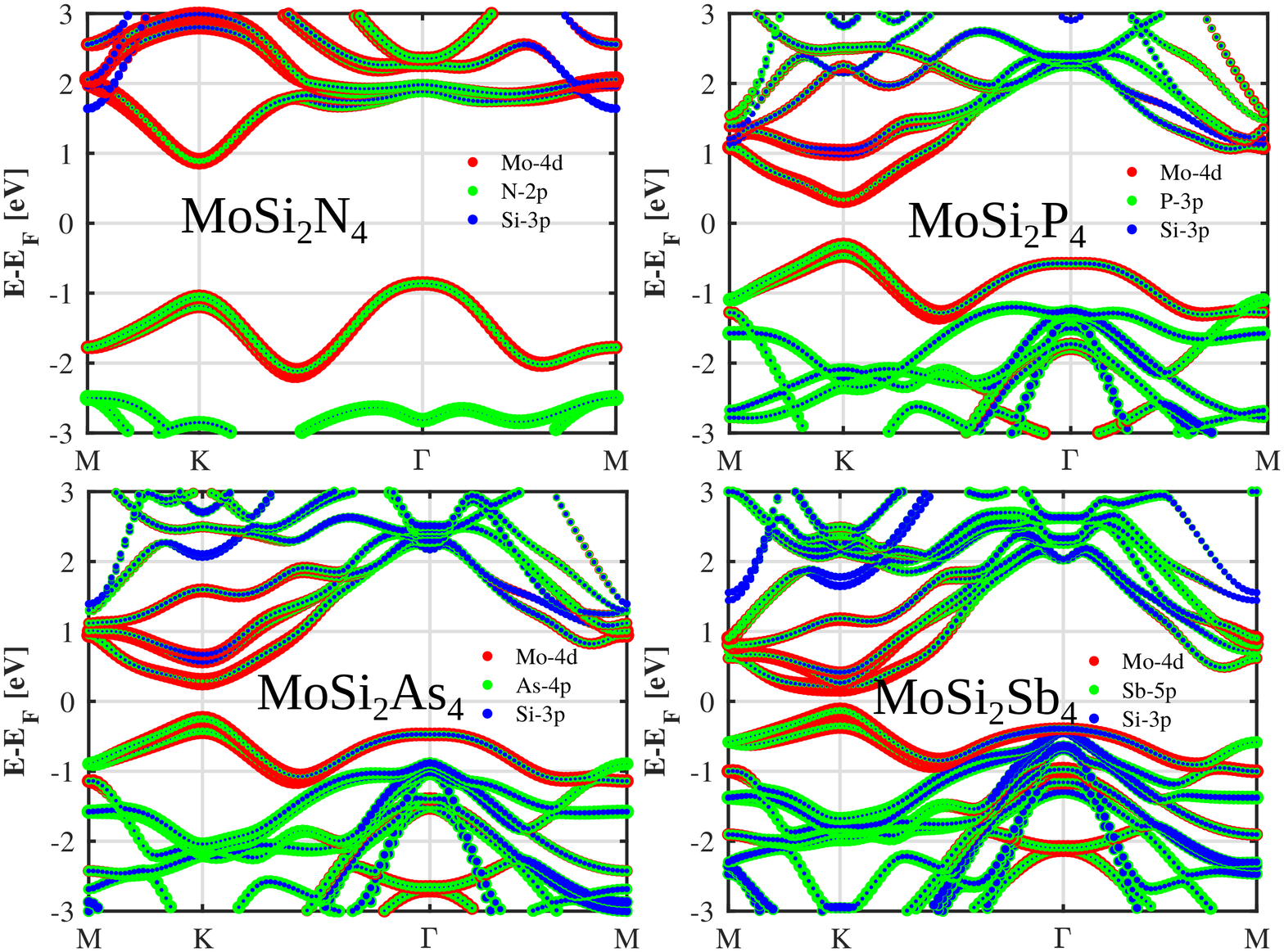}
	\caption{Projected band structures for the studied materials. The d-orbital of Mo along with the p-orbital of Si and X atoms have the main contributions around the band gaps.  }
	\label{fig:orbitalbandstructure}
\end{figure*}

The energies of the K-valleys for the first and second conduction bands, and K- and $\Gamma$-points of the valence band versus vertical strain is plotted in Fig. \ref{fig:energy}. The energies of the up- and down-spins are separately shown in the figure. The strain ranges of the figures are different and the strain ranges are selected in the semiconducting phase. At large compressive strains, the energy of $\Gamma$-point of the valence band is higher than K-point and K-valley determines the VBM. By increasing strain from compressive to the tensile regime, the energy of $\Gamma$-point decreases, and the energy of K-point increases. One can find the large spin-splitting at K-point, whereas $\Gamma$-point does not demonstrate any splitting. On the other hand, two bands contribute to the conduction band. We have called the bands with low and large spin-splitting as $\alpha$-band and $\beta$-band, respectively. The $\alpha$-band contributes as CBM at compressive and small tensile strains. The larger tensile strains bring down the $\beta$-band and change the location of these two bands. Therefore, CBM is contributed by $\beta$-band at a large tensile strain regime.

The charge distribution around K- and $\Gamma$-points of the valence band and K-valleys of the first and second conduction bands in the case of MoSi$_2$P$_4$ is shown in Fig. \ref{fig:charge}. A single layer of MoSi$_2$P$_4$ can be divided into three sub-layers. The charge is significantly localized around the Mo atom in the K-point of valence and conduction bands. $\Gamma$-point of the valence band has bonding character between P and Si atoms of two sub-layers. In addition, the charge is distributed around the most external P atoms. The out-of-plane strain does not significantly affect charge distribution at this point. In the K-point of the valence band, the charge is highly distributed in the internal MoN$_2$ sub-layer and a small charge is around outer P atoms. At compressive out-of-plane strain, the charge around outer P atoms is expanded. Furthermore, the first and second conduction bands at K-valley are compared in the figure. At equilibrium, the charge is localized in the internal sub-layer for the first band, whereas the charge is distributed in the whole thickness of the compound for the second band. The compressive strain does not affect the charge distribution, whereas tensile strain changes the location and charge density of these two bands.

\begin{figure*}
	\centering
	\includegraphics[width=1.0\linewidth]{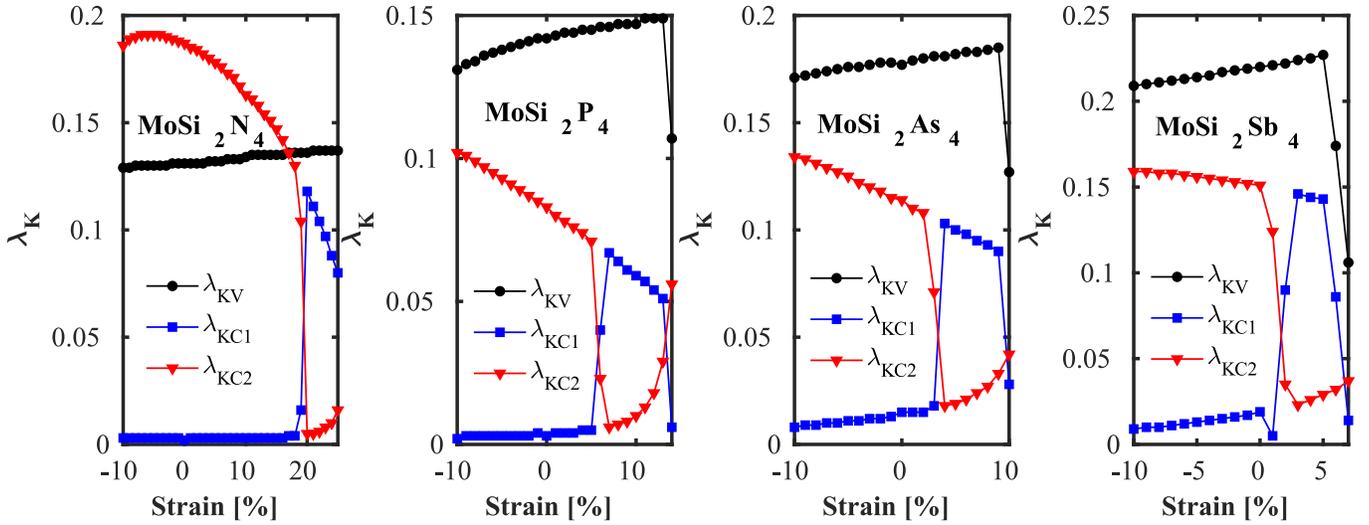}
	\caption{The spin-splitting of the conduction and valence bands at the K-point. The valence band along with the first and second conduction bands are studied. }
	\label{fig:sssb}
\end{figure*}

Two-dimensional maps of the spin-splitting in the first Brillouin zone for all compounds have been shown in Fig. \ref{fig:ssc-ssv}. Both first and second conduction bands along with the valence band are indicated in the figure. The positive and negative spin-splitting around K and K'-valleys are obvious in the figures, respectively. The first conduction bands of all compounds except MoSi$_2$N$_4$ display a low spin-splitting. In the case of MoSi$_2$N$_4$, a large spin-splitting is around M- and \textGamma-valley, but these two valleys do not contribute to the CBM. One can observe that the spin-splitting of the first conduction band is approximately zero on the boundary of the first Brillouin zone for MoSi$_2$P$_4$ and MoSi$_2$As$_4$. On the contrary, the second conduction band exhibits a high spin-splitting which is located at K- and K'- valleys. MoSi$_2$N$_4$ and MoSi$_2$Sb$_4$ demonstrate a higher spin-splitting compared to the others. ُThe valence band shows a tidy spin-splitting so that the splitting is maximum around K- and K'-valleys and it vanishes elsewhere. The spin-splitting increases for heavier compounds and MoSi$_2$Sb$_4$ show the highest value. We have seen from Table \ref{tab:tab2} that the valence band of MoSi$_2$Sb$_4$ has the highest spin-splitting as high as 220 meV. The spin-splitting of the valence band is isotropic around the K-valley of MoSi$_2$P$_4$ and other materials specially MoSi$_2$Sb$_4$ demonstrate anisotropic spin-splitting. In this regard, the contour around the K-valley of MoSi$_2$P$_4$ is circular, whereas MoSi$_2$Sb$_4$ has a triangular shape. 

To clarify the origin of spin-splitting in the valence and conduction bands, the projected band structure of four structures is plotted in Fig. \ref{fig:orbitalbandstructure}. As it is obvious, the first and second conduction bands at K-valley are mainly affected by the d-orbitals of molybdenum atoms. The p-orbitals of X and Si atoms have a low contribution on the first and second bands, respectively. The large spin-splitting of the second band comes from the d-orbitals, whereas the first conduction band with a high d-orbital contribution shows a small splitting. Maybe, this is related to the odd and even bands because the third one also displays a negligible spin-splitting in MoSi$_2$As$_4$ and MoSi$_2$Sb$_4$. In addition, the valence band at K- and \textGamma-points is highly formed by the d-orbitals of Mo atoms. The p-orbitals of X atoms also have a considerable influence at the K-point, whereas p-orbitals of Si atoms contribute to the \textGamma-point. Therefore, the d-orbitals of Mo atoms, the p-orbitals of X and Si atoms have the most impact on the \textGamma-point of the valence band, respectively. The large spin-splitting at valence band is originated from the d-orbitals, whereas the p-orbitals of X atoms enhance the spin-splitting for heavier compounds.  

The spin-splitting of the K-point of the valence band and the first and second bands of the conduction band versus vertical strain is shown in Fig. \ref{fig:sssb}. $\lambda_{K,V}$ increases slightly when strain increases from compressive to tensile regime. The increase of the spin-splitting of the valence band with respect to strain has also observed in the group-III monochalcogenides \cite{ariapour2020strain}. One can find from the band structures and Table \ref{tab:tab2}, that the first conduction band has a negligible spin-splitting, while the second band demonstrates a considerable spin-splitting. $\lambda_{K,C1}$ increases when strain increases from compressive to tensile, whereas $\lambda_{K,C2}$ decreases. The slope of $\lambda_{K,C2}$ variation is higher in lighter compounds. As one can observe the curves of $\lambda_{K,C1}$ and $\lambda_{K,C2}$ change with each other at the tensile strain. As we have seen in Fig. \ref{fig:energy}, the $\beta$-band moves downward and becomes lower than $\alpha$-band for tensile strain. So, in the tensile regime, $\beta$-band with large spin-splitting is the first conduction band. This band changing is also observed in the figure.

%%%%%%%%%%%%%%%%%%%%%%%%%%%%%%%%%%%%%%%%%%%%%%%%%%%%%%%%%%%%%%%%%%%%%%%%%%%%%%%%%%%%%%%
%%%%%%%%%%%%%%%%%%%%%%%%%%%%%%%%  C O N C L U S I O N  %%%%%%%%%%%%%%%%%%%%%%%%%%%%%%%%
%%%%%%%%%%%%%%%%%%%%%%%%%%%%%%%%%%%%%%%%%%%%%%%%%%%%%%%%%%%%%%%%%%%%%%%%%%%%%%%%%%%%%%%
\section{conclusion}
\label{Sec:Conclusions}
The electrical and spin properties of monolayer MoSi$_2$X$_4$ (X = N, P, As, and Sb) under vertical strain are investigated. MoSi$_2$N$_4$ demonstrates an indirect bandgap from $\Gamma$ at valence to K at conduction band, whereas K-valley at valence band is close to $\Gamma$-point and contributes to the valence band. On the other hand, three other compounds have direct bandgap at K-point and $\Gamma$-point at valence band is also close to K-point. Although the first conduction band demonstrates a small spin-splitting, the second conduction band has a large spin-splitting and its minimum is near to the first band.
The PDOS demonstrates that the conduction and valence bands are mostly affected by the d-orbitals of Mo atoms. X atoms have a lower contribution to the valence band. The projected band structure also demonstrates that the conduction and valence bands are mainly constructed from the d-orbitals of Mo atoms. The p-orbitals of X and Si atoms also contribute to the valence and second conduction bands, respectively.
The spin-splitting at K-point ($\lambda_{K,V}$) of valence band increases with changing X-atom from N to Sb and reaches 220 meV for MoSi$_2$Sb$_4$.
The spin-splitting at K-valley in the second conduction band ($\lambda_{K,C2}$) increases from 83 meV in MoSi$_2$P$_4$ to 151 meV for MoSi$_2$Sb$_4$. In the following, the effect of vertical strain has been investigated. The band gaps exhibit a maximum value at small tensile strains, then they decrease for compressive and larger tensile strains. 
The band gaps close at compressive transition strains ($\varepsilon_{trans}$) of around -10$\%$ for all compounds except MoSi$_2$N$_4$ that its bandgap vanishes at the larger strain of about -22$\%$. 
Transition pressures are also distributed from 9.1 to 7.3 GPa for MoSi$_2$P$_4$, MoSi$_2$As$_4$ and MoSi$_2$Sb$_4$, whereas MoSi$_2$N$_4$ displays a larger $P_{trans}$ of 25.3 GPa. In addition, the charge is mostly localized around Mo atoms in both valence and conduction bands. The spin-splitting around K-valley at the valence band is isotropic and circular in the case of MoSi$_2$P$_4$, whereas the anisotropy increases for heavier compounds and MoSi$_2$Sb$_4$ exhibits a triangular shape contour. The spin-splitting of K-valley of the valence band increases when strain increases from compressive to tensile regime. $\lambda_{K,C1}$ increases when strain increases from compressive to tensile, and at the same time, $\lambda_{K,C2}$ decreases. The slope of $\lambda_{K,C2}$ variation is higher for the lighter compounds. $\lambda_{K,C1}$ and $\lambda_{K,C2}$ change with each other at the tensile strain. In addition to the electrical properties, the vertical strain can also control the spin properties.

%%%%%%%%%%%%%%%%%%%%%%%%%%%%%%%%%%%%%%%%%%%%%%%%%%%%%%%%%%%%%%%%%%%%%%%%%%%%%%%%%%%%%%%
%%%%%%%%%%%%%%%%%%%%%%%%%%%%%%%%  R E F E R E N C E S  %%%%%%%%%%%%%%%%%%%%%%%%%%%%%%%%
%%%%%%%%%%%%%%%%%%%%%%%%%%%%%%%%%%%%%%%%%%%%%%%%%%%%%%%%%%%%%%%%%%%%%%%%%%%%%%%%%%%%%%%
\bibliographystyle{apsrev}
\bibliography{acronym,MoSiN,DFT}

\begin{thebibliography}{33}
\expandafter\ifx\csname natexlab\endcsname\relax\def\natexlab#1{#1}\fi
\expandafter\ifx\csname bibnamefont\endcsname\relax
  \def\bibnamefont#1{#1}\fi
\expandafter\ifx\csname bibfnamefont\endcsname\relax
  \def\bibfnamefont#1{#1}\fi
\expandafter\ifx\csname citenamefont\endcsname\relax
  \def\citenamefont#1{#1}\fi
\expandafter\ifx\csname url\endcsname\relax
  \def\url#1{\texttt{#1}}\fi
\expandafter\ifx\csname urlprefix\endcsname\relax\def\urlprefix{URL }\fi
\providecommand{\bibinfo}[2]{#2}
\providecommand{\eprint}[2][]{\url{#2}}

\bibitem[{\citenamefont{Neto et~al.}(2009)\citenamefont{Neto, Guinea, Peres,
  Novoselov, and Geim}}]{neto2009electronic}
\bibinfo{author}{\bibfnamefont{A.~C.} \bibnamefont{Neto}},
  \bibinfo{author}{\bibfnamefont{F.}~\bibnamefont{Guinea}},
  \bibinfo{author}{\bibfnamefont{N.~M.} \bibnamefont{Peres}},
  \bibinfo{author}{\bibfnamefont{K.~S.} \bibnamefont{Novoselov}},
  \bibnamefont{and} \bibinfo{author}{\bibfnamefont{A.~K.} \bibnamefont{Geim}},
  \bibinfo{journal}{Reviews of modern physics} \textbf{\bibinfo{volume}{81}},
  \bibinfo{pages}{109} (\bibinfo{year}{2009}).

\bibitem[{\citenamefont{Radisavljevic et~al.}(2011)\citenamefont{Radisavljevic,
  Radenovic, Brivio, Giacometti, and Kis}}]{radisavljevic2011single}
\bibinfo{author}{\bibfnamefont{B.}~\bibnamefont{Radisavljevic}},
  \bibinfo{author}{\bibfnamefont{A.}~\bibnamefont{Radenovic}},
  \bibinfo{author}{\bibfnamefont{J.}~\bibnamefont{Brivio}},
  \bibinfo{author}{\bibfnamefont{V.}~\bibnamefont{Giacometti}},
  \bibnamefont{and} \bibinfo{author}{\bibfnamefont{A.}~\bibnamefont{Kis}},
  \bibinfo{journal}{Nature nanotechnology} \textbf{\bibinfo{volume}{6}},
  \bibinfo{pages}{147} (\bibinfo{year}{2011}).

\bibitem[{\citenamefont{Guan et~al.}(2014)\citenamefont{Guan, Zhu, and
  Tom{\'a}nek}}]{guan2014phase}
\bibinfo{author}{\bibfnamefont{J.}~\bibnamefont{Guan}},
  \bibinfo{author}{\bibfnamefont{Z.}~\bibnamefont{Zhu}}, \bibnamefont{and}
  \bibinfo{author}{\bibfnamefont{D.}~\bibnamefont{Tom{\'a}nek}},
  \bibinfo{journal}{Physical review letters} \textbf{\bibinfo{volume}{113}},
  \bibinfo{pages}{046804} (\bibinfo{year}{2014}).

\bibitem[{\citenamefont{Zhang et~al.}(2015)\citenamefont{Zhang, Yan, Li, Chen,
  and Zeng}}]{zhang2015atomically}
\bibinfo{author}{\bibfnamefont{S.}~\bibnamefont{Zhang}},
  \bibinfo{author}{\bibfnamefont{Z.}~\bibnamefont{Yan}},
  \bibinfo{author}{\bibfnamefont{Y.}~\bibnamefont{Li}},
  \bibinfo{author}{\bibfnamefont{Z.}~\bibnamefont{Chen}}, \bibnamefont{and}
  \bibinfo{author}{\bibfnamefont{H.}~\bibnamefont{Zeng}},
  \bibinfo{journal}{Angewandte Chemie} \textbf{\bibinfo{volume}{127}},
  \bibinfo{pages}{3155} (\bibinfo{year}{2015}).

\bibitem[{\citenamefont{Mudd et~al.}(2013)\citenamefont{Mudd, Svatek, Ren,
  Patan{\`e}, Makarovsky, Eaves, Beton, Kovalyuk, Lashkarev, Kudrynskyi
  et~al.}}]{mudd2013tuning}
\bibinfo{author}{\bibfnamefont{G.~W.} \bibnamefont{Mudd}},
  \bibinfo{author}{\bibfnamefont{S.~A.} \bibnamefont{Svatek}},
  \bibinfo{author}{\bibfnamefont{T.}~\bibnamefont{Ren}},
  \bibinfo{author}{\bibfnamefont{A.}~\bibnamefont{Patan{\`e}}},
  \bibinfo{author}{\bibfnamefont{O.}~\bibnamefont{Makarovsky}},
  \bibinfo{author}{\bibfnamefont{L.}~\bibnamefont{Eaves}},
  \bibinfo{author}{\bibfnamefont{P.~H.} \bibnamefont{Beton}},
  \bibinfo{author}{\bibfnamefont{Z.~D.} \bibnamefont{Kovalyuk}},
  \bibinfo{author}{\bibfnamefont{G.~V.} \bibnamefont{Lashkarev}},
  \bibinfo{author}{\bibfnamefont{Z.~R.} \bibnamefont{Kudrynskyi}},
  \bibnamefont{et~al.}, \bibinfo{journal}{Advanced Materials}
  \textbf{\bibinfo{volume}{25}}, \bibinfo{pages}{5714} (\bibinfo{year}{2013}).

\bibitem[{\citenamefont{Li et~al.}(2017)\citenamefont{Li, He, Sun, and
  Zhang}}]{li2017computational}
\bibinfo{author}{\bibfnamefont{Q.}~\bibnamefont{Li}},
  \bibinfo{author}{\bibfnamefont{L.}~\bibnamefont{He}},
  \bibinfo{author}{\bibfnamefont{C.}~\bibnamefont{Sun}}, \bibnamefont{and}
  \bibinfo{author}{\bibfnamefont{X.}~\bibnamefont{Zhang}},
  \bibinfo{journal}{The Journal of Physical Chemistry C}
  \textbf{\bibinfo{volume}{121}}, \bibinfo{pages}{27563}
  (\bibinfo{year}{2017}).

\bibitem[{\citenamefont{Wang et~al.}(2016)\citenamefont{Wang, Wang, Lu, Jiang,
  and Yang}}]{wang2016strain}
\bibinfo{author}{\bibfnamefont{Y.}~\bibnamefont{Wang}},
  \bibinfo{author}{\bibfnamefont{S.-S.} \bibnamefont{Wang}},
  \bibinfo{author}{\bibfnamefont{Y.}~\bibnamefont{Lu}},
  \bibinfo{author}{\bibfnamefont{J.}~\bibnamefont{Jiang}}, \bibnamefont{and}
  \bibinfo{author}{\bibfnamefont{S.~A.} \bibnamefont{Yang}},
  \bibinfo{journal}{Nano letters} \textbf{\bibinfo{volume}{16}},
  \bibinfo{pages}{4576} (\bibinfo{year}{2016}).

\bibitem[{\citenamefont{Ozdemir}(2018)}]{ozdemir2018intercalation}
\bibinfo{author}{\bibfnamefont{B.}~\bibnamefont{Ozdemir}},
  \bibinfo{journal}{Computational Condensed Matter}
  \textbf{\bibinfo{volume}{17}}, \bibinfo{pages}{e00335}
  (\bibinfo{year}{2018}).

\bibitem[{\citenamefont{Hong et~al.}(2020)\citenamefont{Hong, Liu, Wang, Zhou,
  Ma, Xu, Feng, Chen, Chen, Sun et~al.}}]{hong2020chemical}
\bibinfo{author}{\bibfnamefont{Y.-L.} \bibnamefont{Hong}},
  \bibinfo{author}{\bibfnamefont{Z.}~\bibnamefont{Liu}},
  \bibinfo{author}{\bibfnamefont{L.}~\bibnamefont{Wang}},
  \bibinfo{author}{\bibfnamefont{T.}~\bibnamefont{Zhou}},
  \bibinfo{author}{\bibfnamefont{W.}~\bibnamefont{Ma}},
  \bibinfo{author}{\bibfnamefont{C.}~\bibnamefont{Xu}},
  \bibinfo{author}{\bibfnamefont{S.}~\bibnamefont{Feng}},
  \bibinfo{author}{\bibfnamefont{L.}~\bibnamefont{Chen}},
  \bibinfo{author}{\bibfnamefont{M.-L.} \bibnamefont{Chen}},
  \bibinfo{author}{\bibfnamefont{D.-M.} \bibnamefont{Sun}},
  \bibnamefont{et~al.}, \bibinfo{journal}{Science}
  \textbf{\bibinfo{volume}{369}}, \bibinfo{pages}{670} (\bibinfo{year}{2020}).

\bibitem[{\citenamefont{Wang et~al.}(2021)\citenamefont{Wang, Shi, Liu, Zhang,
  Hong, Li, Gao, Chen, Ren, Cheng et~al.}}]{wang2021}
\bibinfo{author}{\bibfnamefont{L.}~\bibnamefont{Wang}},
  \bibinfo{author}{\bibfnamefont{Y.}~\bibnamefont{Shi}},
  \bibinfo{author}{\bibfnamefont{M.}~\bibnamefont{Liu}},
  \bibinfo{author}{\bibfnamefont{A.}~\bibnamefont{Zhang}},
  \bibinfo{author}{\bibfnamefont{Y.}~\bibnamefont{Hong}},
  \bibinfo{author}{\bibfnamefont{R.}~\bibnamefont{Li}},
  \bibinfo{author}{\bibfnamefont{Q.}~\bibnamefont{Gao}},
  \bibinfo{author}{\bibfnamefont{M.}~\bibnamefont{Chen}},
  \bibinfo{author}{\bibfnamefont{W.}~\bibnamefont{Ren}},
  \bibinfo{author}{\bibfnamefont{H.}~\bibnamefont{Cheng}},
  \bibnamefont{et~al.}, \bibinfo{journal}{Nature Communications}
  \textbf{\bibinfo{volume}{12}} (\bibinfo{year}{2021}).

\bibitem[{\citenamefont{Zhong et~al.}(2021)\citenamefont{Zhong, Xiong, Lv, Yu,
  and Yuan}}]{zhong2021strain}
\bibinfo{author}{\bibfnamefont{H.}~\bibnamefont{Zhong}},
  \bibinfo{author}{\bibfnamefont{W.}~\bibnamefont{Xiong}},
  \bibinfo{author}{\bibfnamefont{P.}~\bibnamefont{Lv}},
  \bibinfo{author}{\bibfnamefont{J.}~\bibnamefont{Yu}}, \bibnamefont{and}
  \bibinfo{author}{\bibfnamefont{S.}~\bibnamefont{Yuan}},
  \bibinfo{journal}{Physical Review B} \textbf{\bibinfo{volume}{103}},
  \bibinfo{pages}{085124} (\bibinfo{year}{2021}).

\bibitem[{\citenamefont{Li et~al.}(2021)\citenamefont{Li, Geng, Ai, Bai, Lo,
  Ng, Kawazoe, Pan et~al.}}]{li2021design}
\bibinfo{author}{\bibfnamefont{B.}~\bibnamefont{Li}},
  \bibinfo{author}{\bibfnamefont{J.}~\bibnamefont{Geng}},
  \bibinfo{author}{\bibfnamefont{H.}~\bibnamefont{Ai}},
  \bibinfo{author}{\bibfnamefont{H.}~\bibnamefont{Bai}},
  \bibinfo{author}{\bibfnamefont{K.~H.} \bibnamefont{Lo}},
  \bibinfo{author}{\bibfnamefont{K.~W.} \bibnamefont{Ng}},
  \bibinfo{author}{\bibfnamefont{Y.}~\bibnamefont{Kawazoe}},
  \bibinfo{author}{\bibfnamefont{H.}~\bibnamefont{Pan}}, \bibnamefont{et~al.},
  \bibinfo{journal}{Nanoscale}  (\bibinfo{year}{2021}).

\bibitem[{\citenamefont{Mortazavi et~al.}(2021)\citenamefont{Mortazavi,
  Javvaji, Shojaei, Rabczuk, Shapeev, and Zhuang}}]{mortazavi2021exceptional}
\bibinfo{author}{\bibfnamefont{B.}~\bibnamefont{Mortazavi}},
  \bibinfo{author}{\bibfnamefont{B.}~\bibnamefont{Javvaji}},
  \bibinfo{author}{\bibfnamefont{F.}~\bibnamefont{Shojaei}},
  \bibinfo{author}{\bibfnamefont{T.}~\bibnamefont{Rabczuk}},
  \bibinfo{author}{\bibfnamefont{A.~V.} \bibnamefont{Shapeev}},
  \bibnamefont{and} \bibinfo{author}{\bibfnamefont{X.}~\bibnamefont{Zhuang}},
  \bibinfo{journal}{Nano Energy} \textbf{\bibinfo{volume}{82}},
  \bibinfo{pages}{105716} (\bibinfo{year}{2021}).

\bibitem[{\citenamefont{Yu et~al.}(2021)\citenamefont{Yu, Zhou, Wan, and
  Li}}]{yu2021high}
\bibinfo{author}{\bibfnamefont{J.}~\bibnamefont{Yu}},
  \bibinfo{author}{\bibfnamefont{J.}~\bibnamefont{Zhou}},
  \bibinfo{author}{\bibfnamefont{X.}~\bibnamefont{Wan}}, \bibnamefont{and}
  \bibinfo{author}{\bibfnamefont{Q.}~\bibnamefont{Li}}, \bibinfo{journal}{New
  Journal of Physics} \textbf{\bibinfo{volume}{23}}, \bibinfo{pages}{033005}
  (\bibinfo{year}{2021}).

\bibitem[{\citenamefont{Bafekry
  et~al.}(2021{\natexlab{a}})\citenamefont{Bafekry, Faraji, Hoat, Shahrokhi,
  Fadlallah, Shojaei, Feghhi, Ghergherehchi, and Gogova}}]{bafekry2021mosi2n4}
\bibinfo{author}{\bibfnamefont{A.}~\bibnamefont{Bafekry}},
  \bibinfo{author}{\bibfnamefont{M.}~\bibnamefont{Faraji}},
  \bibinfo{author}{\bibfnamefont{D.}~\bibnamefont{Hoat}},
  \bibinfo{author}{\bibfnamefont{M.}~\bibnamefont{Shahrokhi}},
  \bibinfo{author}{\bibfnamefont{M.}~\bibnamefont{Fadlallah}},
  \bibinfo{author}{\bibfnamefont{F.}~\bibnamefont{Shojaei}},
  \bibinfo{author}{\bibfnamefont{S.~A.~H.} \bibnamefont{Feghhi}},
  \bibinfo{author}{\bibfnamefont{M.}~\bibnamefont{Ghergherehchi}},
  \bibnamefont{and} \bibinfo{author}{\bibfnamefont{D.}~\bibnamefont{Gogova}},
  \bibinfo{journal}{Journal of Physics D: Applied Physics}
  \textbf{\bibinfo{volume}{54}}, \bibinfo{pages}{155303}
  (\bibinfo{year}{2021}{\natexlab{a}}).

\bibitem[{\citenamefont{Yao et~al.}(2021)\citenamefont{Yao, Zhang, Wang, Li,
  Yu, Xu, Wang, and Wei}}]{yao2021novel}
\bibinfo{author}{\bibfnamefont{H.}~\bibnamefont{Yao}},
  \bibinfo{author}{\bibfnamefont{C.}~\bibnamefont{Zhang}},
  \bibinfo{author}{\bibfnamefont{Q.}~\bibnamefont{Wang}},
  \bibinfo{author}{\bibfnamefont{J.}~\bibnamefont{Li}},
  \bibinfo{author}{\bibfnamefont{Y.}~\bibnamefont{Yu}},
  \bibinfo{author}{\bibfnamefont{F.}~\bibnamefont{Xu}},
  \bibinfo{author}{\bibfnamefont{B.}~\bibnamefont{Wang}}, \bibnamefont{and}
  \bibinfo{author}{\bibfnamefont{Y.}~\bibnamefont{Wei}},
  \bibinfo{journal}{Nanomaterials} \textbf{\bibinfo{volume}{11}},
  \bibinfo{pages}{559} (\bibinfo{year}{2021}).

\bibitem[{\citenamefont{Yang et~al.}(2021)\citenamefont{Yang, Song, Sun, and
  Lu}}]{yang2021valley}
\bibinfo{author}{\bibfnamefont{C.}~\bibnamefont{Yang}},
  \bibinfo{author}{\bibfnamefont{Z.}~\bibnamefont{Song}},
  \bibinfo{author}{\bibfnamefont{X.}~\bibnamefont{Sun}}, \bibnamefont{and}
  \bibinfo{author}{\bibfnamefont{J.}~\bibnamefont{Lu}},
  \bibinfo{journal}{Physical Review B} \textbf{\bibinfo{volume}{103}},
  \bibinfo{pages}{035308} (\bibinfo{year}{2021}).

\bibitem[{\citenamefont{Ai et~al.}(2021)\citenamefont{Ai, Liu, Geng, Wang, Lo,
  and Pan}}]{ai2021theoretical}
\bibinfo{author}{\bibfnamefont{H.}~\bibnamefont{Ai}},
  \bibinfo{author}{\bibfnamefont{D.}~\bibnamefont{Liu}},
  \bibinfo{author}{\bibfnamefont{J.}~\bibnamefont{Geng}},
  \bibinfo{author}{\bibfnamefont{S.}~\bibnamefont{Wang}},
  \bibinfo{author}{\bibfnamefont{K.~H.} \bibnamefont{Lo}}, \bibnamefont{and}
  \bibinfo{author}{\bibfnamefont{H.}~\bibnamefont{Pan}},
  \bibinfo{journal}{Physical Chemistry Chemical Physics}
  \textbf{\bibinfo{volume}{23}}, \bibinfo{pages}{3144} (\bibinfo{year}{2021}).

\bibitem[{\citenamefont{Nayak et~al.}(2014)\citenamefont{Nayak, Bhattacharyya,
  Zhu, Liu, Wu, Pandey, Jin, Singh, Akinwande, and Lin}}]{nayak2014pressure}
\bibinfo{author}{\bibfnamefont{A.~P.} \bibnamefont{Nayak}},
  \bibinfo{author}{\bibfnamefont{S.}~\bibnamefont{Bhattacharyya}},
  \bibinfo{author}{\bibfnamefont{J.}~\bibnamefont{Zhu}},
  \bibinfo{author}{\bibfnamefont{J.}~\bibnamefont{Liu}},
  \bibinfo{author}{\bibfnamefont{X.}~\bibnamefont{Wu}},
  \bibinfo{author}{\bibfnamefont{T.}~\bibnamefont{Pandey}},
  \bibinfo{author}{\bibfnamefont{C.}~\bibnamefont{Jin}},
  \bibinfo{author}{\bibfnamefont{A.~K.} \bibnamefont{Singh}},
  \bibinfo{author}{\bibfnamefont{D.}~\bibnamefont{Akinwande}},
  \bibnamefont{and} \bibinfo{author}{\bibfnamefont{J.-F.} \bibnamefont{Lin}},
  \bibinfo{journal}{Nature communications} \textbf{\bibinfo{volume}{5}},
  \bibinfo{pages}{1} (\bibinfo{year}{2014}).

\bibitem[{\citenamefont{Chi et~al.}(2014)\citenamefont{Chi, Zhao, Zhang,
  Goncharov, Lobanov, Kagayama, Sakata, and Chen}}]{chi2014pressure}
\bibinfo{author}{\bibfnamefont{Z.-H.} \bibnamefont{Chi}},
  \bibinfo{author}{\bibfnamefont{X.-M.} \bibnamefont{Zhao}},
  \bibinfo{author}{\bibfnamefont{H.}~\bibnamefont{Zhang}},
  \bibinfo{author}{\bibfnamefont{A.~F.} \bibnamefont{Goncharov}},
  \bibinfo{author}{\bibfnamefont{S.~S.} \bibnamefont{Lobanov}},
  \bibinfo{author}{\bibfnamefont{T.}~\bibnamefont{Kagayama}},
  \bibinfo{author}{\bibfnamefont{M.}~\bibnamefont{Sakata}}, \bibnamefont{and}
  \bibinfo{author}{\bibfnamefont{X.-J.} \bibnamefont{Chen}},
  \bibinfo{journal}{Physical review letters} \textbf{\bibinfo{volume}{113}},
  \bibinfo{pages}{036802} (\bibinfo{year}{2014}).

\bibitem[{\citenamefont{Ghobadi}(2019)}]{ghobadi2019normal}
\bibinfo{author}{\bibfnamefont{N.}~\bibnamefont{Ghobadi}},
  \bibinfo{journal}{Physica E} \textbf{\bibinfo{volume}{111}},
  \bibinfo{pages}{158} (\bibinfo{year}{2019}).

\bibitem[{\citenamefont{Bafekry
  et~al.}(2021{\natexlab{b}})\citenamefont{Bafekry, Stampfl, Naseri, Fadlallah,
  Faraji, Ghergherehchi, Gogova, and Feghhi}}]{bafekry2021effect}
\bibinfo{author}{\bibfnamefont{A.}~\bibnamefont{Bafekry}},
  \bibinfo{author}{\bibfnamefont{C.}~\bibnamefont{Stampfl}},
  \bibinfo{author}{\bibfnamefont{M.}~\bibnamefont{Naseri}},
  \bibinfo{author}{\bibfnamefont{M.~M.} \bibnamefont{Fadlallah}},
  \bibinfo{author}{\bibfnamefont{M.}~\bibnamefont{Faraji}},
  \bibinfo{author}{\bibfnamefont{M.}~\bibnamefont{Ghergherehchi}},
  \bibinfo{author}{\bibfnamefont{D.}~\bibnamefont{Gogova}}, \bibnamefont{and}
  \bibinfo{author}{\bibfnamefont{S.}~\bibnamefont{Feghhi}},
  \bibinfo{journal}{Journal of Applied Physics} \textbf{\bibinfo{volume}{129}},
  \bibinfo{pages}{155103} (\bibinfo{year}{2021}{\natexlab{b}}).

\bibitem[{\citenamefont{Wu et~al.}(2021)\citenamefont{Wu, Cao, Ang, and
  Ang}}]{wu2021semiconductor}
\bibinfo{author}{\bibfnamefont{Q.}~\bibnamefont{Wu}},
  \bibinfo{author}{\bibfnamefont{L.}~\bibnamefont{Cao}},
  \bibinfo{author}{\bibfnamefont{Y.~S.} \bibnamefont{Ang}}, \bibnamefont{and}
  \bibinfo{author}{\bibfnamefont{L.~K.} \bibnamefont{Ang}},
  \bibinfo{journal}{Applied Physics Letters} \textbf{\bibinfo{volume}{118}},
  \bibinfo{pages}{113102} (\bibinfo{year}{2021}).

\bibitem[{\citenamefont{Guo et~al.}(2020)\citenamefont{Guo, Mu, Zhu, and
  Chen}}]{guo2020coexistence}
\bibinfo{author}{\bibfnamefont{S.-D.} \bibnamefont{Guo}},
  \bibinfo{author}{\bibfnamefont{W.-Q.} \bibnamefont{Mu}},
  \bibinfo{author}{\bibfnamefont{Y.-T.} \bibnamefont{Zhu}}, \bibnamefont{and}
  \bibinfo{author}{\bibfnamefont{X.-Q.} \bibnamefont{Chen}},
  \bibinfo{journal}{Physical Chemistry Chemical Physics}
  \textbf{\bibinfo{volume}{22}}, \bibinfo{pages}{28359} (\bibinfo{year}{2020}).

\bibitem[{\citenamefont{Soler et~al.}(2002)\citenamefont{Soler, Artacho, Gale,
  Garc{\'\i}a, Junquera, Ordej{\'o}n, and
  S{\'a}nchez-Portal}}]{soler2002siesta}
\bibinfo{author}{\bibfnamefont{J.~M.} \bibnamefont{Soler}},
  \bibinfo{author}{\bibfnamefont{E.}~\bibnamefont{Artacho}},
  \bibinfo{author}{\bibfnamefont{J.~D.} \bibnamefont{Gale}},
  \bibinfo{author}{\bibfnamefont{A.}~\bibnamefont{Garc{\'\i}a}},
  \bibinfo{author}{\bibfnamefont{J.}~\bibnamefont{Junquera}},
  \bibinfo{author}{\bibfnamefont{P.}~\bibnamefont{Ordej{\'o}n}},
  \bibnamefont{and}
  \bibinfo{author}{\bibfnamefont{D.}~\bibnamefont{S{\'a}nchez-Portal}},
  \bibinfo{journal}{J. Phys.:Condensed Matter} \textbf{\bibinfo{volume}{14}},
  \bibinfo{pages}{2745} (\bibinfo{year}{2002}).

\bibitem[{\citenamefont{Perdew and Zunger}(1981)}]{perdew1981self}
\bibinfo{author}{\bibfnamefont{J.~P.} \bibnamefont{Perdew}} \bibnamefont{and}
  \bibinfo{author}{\bibfnamefont{A.}~\bibnamefont{Zunger}},
  \bibinfo{journal}{Phys. Rev. B} \textbf{\bibinfo{volume}{23}},
  \bibinfo{pages}{5048} (\bibinfo{year}{1981}).

\bibitem[{\citenamefont{Kokalj}(2003)}]{kokalj2003computer}
\bibinfo{author}{\bibfnamefont{A.}~\bibnamefont{Kokalj}},
  \textbf{\bibinfo{volume}{28}}, \bibinfo{pages}{155} (\bibinfo{year}{2003}).

\bibitem[{\citenamefont{Touski and Ghobadi}(2020)}]{touski2020interplay}
\bibinfo{author}{\bibfnamefont{S.~B.} \bibnamefont{Touski}} \bibnamefont{and}
  \bibinfo{author}{\bibfnamefont{N.}~\bibnamefont{Ghobadi}},
  \bibinfo{journal}{Physica E} p. \bibinfo{pages}{114407}
  (\bibinfo{year}{2020}).

\bibitem[{\citenamefont{Ghobadi and Touski}(2020)}]{ghobadi2020electrical}
\bibinfo{author}{\bibfnamefont{N.}~\bibnamefont{Ghobadi}} \bibnamefont{and}
  \bibinfo{author}{\bibfnamefont{S.~B.} \bibnamefont{Touski}},
  \bibinfo{journal}{Journal of Physics: Condensed Matter}
  \textbf{\bibinfo{volume}{33}}, \bibinfo{pages}{085502}
  (\bibinfo{year}{2020}).

\bibitem[{\citenamefont{Li et~al.}(2020)\citenamefont{Li, Wu, Feng, Guan, Feng,
  Yao, and Yang}}]{li2020valley}
\bibinfo{author}{\bibfnamefont{S.}~\bibnamefont{Li}},
  \bibinfo{author}{\bibfnamefont{W.}~\bibnamefont{Wu}},
  \bibinfo{author}{\bibfnamefont{X.}~\bibnamefont{Feng}},
  \bibinfo{author}{\bibfnamefont{S.}~\bibnamefont{Guan}},
  \bibinfo{author}{\bibfnamefont{W.}~\bibnamefont{Feng}},
  \bibinfo{author}{\bibfnamefont{Y.}~\bibnamefont{Yao}}, \bibnamefont{and}
  \bibinfo{author}{\bibfnamefont{S.~A.} \bibnamefont{Yang}},
  \bibinfo{journal}{Physical Review B} \textbf{\bibinfo{volume}{102}},
  \bibinfo{pages}{235435} (\bibinfo{year}{2020}).

\bibitem[{\citenamefont{Ghobadi and Touski}(2021)}]{ghobadi2021structural}
\bibinfo{author}{\bibfnamefont{N.}~\bibnamefont{Ghobadi}} \bibnamefont{and}
  \bibinfo{author}{\bibfnamefont{S.~B.} \bibnamefont{Touski}},
  \bibinfo{journal}{Journal of Physics: Condensed Matter}
  (\bibinfo{year}{2021}).

\bibitem[{\citenamefont{Shamekhi and Ghobadi}(2020)}]{shamekhi2020band}
\bibinfo{author}{\bibfnamefont{M.}~\bibnamefont{Shamekhi}} \bibnamefont{and}
  \bibinfo{author}{\bibfnamefont{N.}~\bibnamefont{Ghobadi}},
  \bibinfo{journal}{Physica B} \textbf{\bibinfo{volume}{580}},
  \bibinfo{pages}{411923} (\bibinfo{year}{2020}).

\bibitem[{\citenamefont{Ariapour and Touski}(2020)}]{ariapour2020strain}
\bibinfo{author}{\bibfnamefont{M.}~\bibnamefont{Ariapour}} \bibnamefont{and}
  \bibinfo{author}{\bibfnamefont{S.~B.} \bibnamefont{Touski}},
  \bibinfo{journal}{Journal of Magnetism and Magnetic Materials}
  \textbf{\bibinfo{volume}{510}}, \bibinfo{pages}{166922}
  (\bibinfo{year}{2020}).

\end{thebibliography}

\end{document}